\documentclass[article,pra,longbibliography]{revtex4-1}  

\usepackage{amsthm}
\usepackage{amsmath}    
\usepackage{amsfonts} 
\usepackage{amssymb}
\usepackage{graphicx}   
\usepackage{algorithm}
\usepackage[colorlinks,breaklinks=true]{hyperref}
\usepackage{algpseudocode}
\usepackage{mathtools}
\DeclarePairedDelimiter\floor{\lfloor}{\rfloor}
\DeclarePairedDelimiter\ceil{\lceil}{\rceil}
\theoremstyle{break}

\newtheorem{thm}{Theorem}
\newtheorem{definition}{Definition} 
\newtheorem{corollary}{Corollary}[thm] 

\newcommand{\ket}[1]{\left|#1\right\rangle}

\newcommand{\1}{\ket{1}}

\newcommand{\ord}{\operatorname{ord}}
\newcommand{\lcm}{\operatorname{lcm}}

\begin{document}

\title{A quantum algorithm for computing the Carmichael function}
\author{Juan Carlos Garcia-Escartin}
\email{juagar@tel.uva.es}  
\affiliation{Universidad de Valladolid, Dpto. Teor\'ia de la Se\~{n}al e Ing. Telem\'atica, Paseo Bel\'en n$^o$ 15, 47011 Valladolid, Spain}
\date{\today}

\begin{abstract}
Quantum computers can solve many number theory problems efficiently. Using the efficient quantum algorithm for order finding as an oracle, this paper presents an algorithm that computes the Carmichael function for any integer $N$ with a probability as close to 1 as desired. The algorithm requires $O((\log n )^3n^3)$ quantum operations, or $O(\log\log n (\log n)^4 n^2)$ operations using fast multiplication. Verification, quantum optimizations and applications to RSA and primality tests are also discussed.
\end{abstract}
\maketitle
\section{Introduction}
In problems with a strong inner structure, quantum computers can provide efficient algorithms which outperform any known classical method. In particular, many problems in number theory have an efficient quantum solution. Some examples are factoring integers and the discrete logarithm problem \cite{Sho97}, determining whether a number is square-free or not \cite{LPS12} and finding hidden shifts in multiplicative characters \cite{vDHI06}.

This paper presents a method to compute the Carmichael function for any integer $N$ using a quantum order finding subroutine. The method generalizes a previous quantum primality test \cite{DG18} and can also be used to determine primality or compositeness with high probability. 

\section{Definitions and notation}
We will work with the \emph{multiplicative group of integers modulo N} defined as $\mathbb{Z}_N^*=\{a \in \mathbb{Z}_N : (a,N)=1\}$ where $\mathbb{Z}_N$ is the ring of integers modulo $N$ and $(a,N)$ the greatest common divisor of $a$ and $N$. The elements of $\mathbb{Z}_N^*$ are the integers from $1$ to $N-1$ which are coprime to $N$. These integers form a group under modulo $N$ multiplication. 

For a finite group $G$, the cardinality (number of elements) of $G$ is called the order of the group $|G|$. The order of $\mathbb{Z}_N^*$ is given by Euler's totient function $\varphi (N)$, which is the number of elements $1 \leq a < N$ which are coprime to $N$.

The multiplicative order of an element $a \in \mathbb{Z}_N^*$, $ord_N(a)$, is the smallest positive integer $r$ such that $a^r\equiv 1 \mod N$. For simplicity, $\ord(a)$ will be used as a shorthand for $ord_N(a)$ when the group is $\mathbb{Z}_N^*$ and there is no ambiguity. 

All the elements $a \in \mathbb{Z}_N^*$ so that $a^k\equiv 1 \mod N$ for any constant integer $k$ form a subgroup of $\mathbb{Z}_N^*$ under multiplication. The product of two elements of the group $a_1$ and $a_2$ also has the power property $a_1^k a_2^k \equiv 1 \mod N$ and it remains in the group and each element $a$ has an inverse $a^{k-1}$, which is also in the subgroup. These subgroups contain all the elements of an order which is a divisor of $k$, denoted as $\ord(a)|k$.

The structure of these groups can be studied using two theorems:

\begin{thm}[Lagrange]\label{LagrangeTh}
Let $|G|$ be the number of elements of a finite group $G$, then any subgroup $S$ of $G$ must have a number of elements $|S|$ which is a divisor of the size of the group.
\end{thm}

\begin{thm}[\cite{Bur05}]\label{OrderDiv}
\label{orderdiv}
Let $a \in \mathbb{Z}_N^*$ have order $\ord(a)$. Then $a^h\equiv 1\mod N$ if and only if $\ord(a)| h$.
\end{thm}

Related to the order we also have:
\begin{thm}[Euler]\label{EulerTh}
Let $N$ be a positive integer, then 
$a^{\varphi (N)}\equiv 1 \mod N$
for any positive integer $a$ such that $(a,N)=1$.
\end{thm}
For a prime $N$, $\varphi(N)=N-1$ and we recover Fermat's theorem. 

This paper deals with the {\bf Carmichael function} $\lambda(N)$ of an integer $N$.

\begin{definition}[Carmichael function]\label{CarmichaelDef}
Let $N$ be a positive integer, the Carmichael function $\lambda(N)$ is the smallest integer $r$ such that  
$a^{r}\equiv 1 \mod N$
for all the positive integers $a$ such that $(a,N)=1$.
\end{definition}

A few relevant properties of the Carmichael function are \cite{Car10}:
\begin{itemize}
\item $\lambda(N)$ is the exponent of the group $\mathbb{Z}_N^*$.
\item From Euler and Lagrange theorems, it follows $\lambda(N)| \varphi(N)$. The minimum exponent must be a divisor of $\varphi(N)$.
\item For coprime integers $N_1$ and $N_2$, with $(N_1,N_2)=1$,
\begin{equation}
\label{Lcoprime}
\lambda(N_1 \cdot N_2)=\lcm(\lambda(N_1),\lambda(N_2)),
\end{equation}
where $\lcm$ is the least common multiple function.
\item From the factorization of $N=p_1^{\alpha_1}\cdots p_d^{\alpha_de}$,  $\lambda(N)$ can be computed as
\begin{equation}
\label{LambdaFromFactors1}
\lambda(N)=\lcm(\lambda(p_1^{\alpha_1}),\ldots,\lambda(p_d^{\alpha_d})).
\end{equation}
For prime powers
\begin{eqnarray}
\label{LambdaFromFactors2}
\lambda(p^\alpha)&=&\varphi(p^\alpha)=p^{\alpha-1}(p-1) \text{ for odd } p \text{ or p=2, 4}.\\
\label{LambdaFromFactors3}
\lambda(p^\alpha)&=&\frac{1}{2}\varphi(p^\alpha) \text{ for powers of } 2 \text{ with } \alpha>2.
\end{eqnarray}

\end{itemize}

\section{Problem statement}
Given a positive integer $N$, we want an algorithm that returns $\lambda(N)$. We call $n$ to the number of bits needed to express $2^{n-1}\leq N<2^n$ in binary. 

For a classical computer, this is considered a hard problem that can only be solved with a number of operations which grows almost exponentially with the number of bits of the integer. This assumed difficulty is the foundation of the security of the RSA cryptosystem \cite{RSA78}. However, given the factorization of $N$, computing $\lambda(N)$ becomes trivial using Eqs. (\ref{LambdaFromFactors1}-\ref{LambdaFromFactors3}). 

A quantum computer could use Shor's algorithm \cite{Sho97} to factor $N$ and then find $\lambda(N)$. Instead, this paper presents a direct algorithm with a complexity $O((\log n )^3n^3)$, using the usual big O notation with functions over the number of bits. The quantum part of the algorithm is reduced to a quantum order oracle which returns $\ord(a)$ for any $a\in \mathbb{Z}_N^*$. 

The principles of quantum period finding and how it can be used to give an order oracle are discussed in Section \ref{period}. This section can be skipped if the reader is already familiar with quantum period finding or is not interested in the particulars behind the oracle.

Section \ref{algo} describes an algorithm that computes the Carmichael function for any positive integer $N$ in terms of calls to the order finding oracle and analyzes its expected complexity. A classical method to verify the result is given in Section \ref{verification}.

Quantum computers can also help to further optimize the procedure and in the verification of the results. Section \ref{qimprovements} will discuss some efficiency tradeoffs and speed improvements. The paper concludes with a discussion of the possible applications in Section \ref{discussion}.

\section{Quantum order finding}
\label{period}
One of the key properties of quantum computers is their ability to create and maintain superpositions of states and make them intefere in useful ways. A general quantum state of a system with $M$ posible values can be written as superpostion
\begin{equation}
\sum_{x=0}^{M-1}\alpha_x \ket{x},
\end{equation}
where $\ket{x}$ are the states in the set $\{\ket{0},\ldots,\ket{M-1}\}$ and each $x$ corresponds to an index from 0 to $M-1$. Each state has a probability amplitude $\alpha_x$. These probability amplitudes are complex numbers and $|\alpha_x|^2$ gives the probability of finding the state $\ket{x}$ when we measure the superposition. All the probabilities must sum to 1.

A crucial distinctive feature of quantum computers is that different solutions can interfere like waves. In a quantum algorithm like the order finding algorithm, we take advantage of the structure of the problem to achieve a destructive interference for the unwanted states and a constructive interference for the states encoding the desired solution.

Interference between the states comes from quantum evolution operators which are unitary operators preserving the total probability. 

A basic building block in many quantum algorithms is the Quantum Fourier Transform, or $QFT$, defined as 
\begin{equation}
QFT\ket{x}=\frac{1}{\sqrt{M}}\sum_{y=0}^{M-1} e^{i\frac{2\pi xy}{M}}\ket{y},
\end{equation}
which is just a unitary version of the usual Discrete Fourier Transform. It can also be represented with the usual DFT matrix as long as we take the definition with symmetric $\frac{1}{\sqrt{N}}$ factors both in the direct and the inverse transforms.

The quantum order finding algorithm uses the QFT to find the period of a superposition of modular exponentials. 

\subsection{Quantum period finding}
Consider a periodic function with a period $r$ so that $f(x)=f(y)$ if and only if $x\equiv y \mod r$. Quantum computers can find the unknown hidden period $r$ efficiently \cite{Sho97,HH00}.

We take a set of values from 0 to $M-1$. $M$ is chosen so that $M\gg r$. In that respect, we need an upper bound on $r$, even if, by definition, we cannot know its precise value.

The first step is creating a uniform superposition for all the possible input values from 0 to $M-1$
\begin{equation}
\frac{1}{\sqrt{M}}\sum_{x=0}^{M-1}\ket{x}\ket{0}
\end{equation}
where we use an ancilla state $\ket{0}$ large enough to take any of the output values of $f(x)$ (the image of $f$). One way to create this initial superposition is preparing an initial state $\ket{0}\ket{0}$ and computing the QFT over the first register.

In the second step we need a unitary operation related to the function $f(x)$. A general method to obtain the reversible and unitary operations we need for quantum computers is defining the operator:  
\begin{equation}
U_f\ket{x}\ket{y}=\ket{x}\ket{y+f(x) \mod M}.
\end{equation}
The inverse operation replaces modulo $M$ addition by modulo $M$ subtraction. For binary inputs, the XOR operation can play both roles: subtraction and addition.

Once we have the operator, we compute the superposition
\begin{equation}
\label{StartPeriod}
\frac{1}{\sqrt{M}}\sum_{x=0}^{M-1}\ket{x}\ket{f(x)}.
\end{equation}

We now measure the second register. This fixes its value and the state becomes
\begin{equation}
\label{periodsuperposition}
\frac{1}{\sqrt{m}}\sum_{k=0}^{m-1}\ket{x_0+kr}\ket{f(x_0)},
\end{equation}
where $m$ is the number of times $f(x_0)$ is the output for an input $x$ from $0$ to $M-1$. We are not interested in the exact value of $f(x_0)$ and we will discard it in the following steps. We use the notation $f(x_0)$ where $x_0$ is the smallest value for which $f(x)$ takes the value measured in the second register. Depending on the value of $x_0$, $m=\floor{\frac{M}{r}}$ or $m=\ceil{\frac{M}{r}}$. 

After measuring $f(x_0)$ in the ancillary register, the state is reduced to a superposition of all the inputs $x$ for which $f(x)$ takes the same value.

For the periodic functions we are dealing with, the resulting superposition will take the values $x_0+kr$ for an integer $k$. However, we cannot use this result directly to recover $r$. If we measured the first register, the value $x_0+k_1r$, with a fixed value of $k=k_1$, is not enough to find $r$. After measurement, we no longer have access to the alternative possible results and, if we repeated the procedure, we would obtain a different value $f(x_0')$ and a different set of values $x_0'+kr$. 

However, we can use the Quantum Fourier Transform to deduce the value of $r$. We get
\begin{equation}
QFT \frac{1}{\sqrt{m}}\sum_{k=0}^{m-1}\ket{x_0+kr}=\sum_{y=0}^{M-1}\alpha_y\ket{y}
\end{equation}
with
\begin{equation}
\alpha_y=e^{i2\pi \frac{ x_0}{M}}\frac{1}{\sqrt{Mm}}\sum_{k=0}^{m-1}\left(e^{i2\pi \frac{ry}{M}}\right)^k.
\end{equation}

If we measure this superposition, the probability of finding a concrete $y$ is $|\alpha_y|^2$. In the following, we can ignore the phase factor $e^{i2\pi\frac{ x_0}{M}}$ and concentrate on $|\alpha_y|$. There are two cases. If $M$ is an exact multiple of $r$, then $m=M/r$ and 
\begin{equation}
|\alpha_y|=\frac{1}{\sqrt{Mm}}\sum_{k=0}^{m-1}\left(e^{i2\pi \frac{ y}{m}}\right)^k.
\end{equation}

In this geometric sum, the values corresponding to a state $\ket{y}$ with an integer multiple of $m$, $y=lm$, are made of $m$ terms $e^{i2\pi \frac{ y}{m}}=1$ and each from these states appears with a probability
\begin{equation}
|\alpha_y|^2=\left(\frac{1}{\sqrt{mM}}m\right)^2=\frac{M}{m}=\frac{1}{r}. 
\end{equation}

There are exactly $r$ values of $0\leq y<M$ with this property. The total probability must sum to 1 and we can check there is a completely destructive interference for all the elements $\ket{y}$ which are not multiples of $m$.

At each measurement we get a random multiple $lm$ with $0<l\leq r$. We can recover $m$ from the $\gcd$ of multiple results. We only need to find one $l$ for which $\gcd(l,r)=1$, which happens with high probability after a number of measurements of the order of $\log\log(r)$ (see Eq. \eqref{3Mbound}). Once we have $m$, we can compute $r=M/m$. 

If $M/r$ is not an integer, recovering $r$ is more involved. We still have a constructive interference around the integers closest to $M/r$, but now

\begin{equation}
|\alpha_y|=\frac{1}{\sqrt{Mm}}\frac{\left|1-e^{i2\pi \frac{ mry}{M}}\right|}{\left|1-e^{i2\pi \frac{mr}{M}}\right|}=\frac{1}{\sqrt{Mm}}\frac{\left|\sin\left(\pi\frac{ m r y}{M}\right)\right|}{\left|\sin\left(\pi\frac{ r y}{M}\right)\right|}.
\end{equation}

The procedure is similar to the exact case, but the fraction $M/m$ is reconstructed from methods based on continued fractions. The reader can check the detailed description in the existing literature \cite{Sho97, NC00,Mer07,Ger05}.  

\subsection{Computing the order with quantum period finding}
The quantum period finding algorithm can be directly applied to order finding. For any integer $a\in \mathbb{Z}_N^*$, we can define a function $f(x)=a^x \mod N$ in Eq. \eqref{StartPeriod}. The period of this $f(x)$ is $r=\ord_N(a)$.

We set $M=2^{2n}>N^2$ so that there are enough repeated values to deduce the period. As $\ord_N(a)\leq N-1$, we have a bound on the period and we fulfill all the conditions in the period finding algorithm.
 
The are known quantum circuits that perform the required reversible modular exponentiation with a number of elementary quantum gates of the order of $O(n^3 )$ \cite{VBE96,BCD96}. This is usually the bottleneck in Shor's algorithm and similar methods. Compared to the QFT, which has a number of operations which grow quadratically with the number of bits $n$ \cite{Sho97}, modular exponentiation is the most onerous task. 

In principle, the total complexity can be reduced with fast multiplication like the Sch\"onhage-Strassen algorithm \cite{SS71,Zal98}, which reduces the complexity to $O(\log\log n (\log n) n^2)$. However, these methods, while asymptotically faster, only reduce the total complexity for inputs in the order of thousands of bits \cite{vMI05}. 

There are at most $O(\log\log r)$ repetitions of the preparation and measurement procedure in quantum period finding and each of these stages has a leading term due to modular exponentiation. For our $r<N$ we require $O(\log n)$ repetitions. This makes the total complexity of the quantum order finding oracle $O((\log n )n^3)$ quantum operations, or $O(\log\log n (\log n)^2 n^2)$ for fast multiplication.

The number of operations in the classical part is less than the cubic complexity of modular exponentiation. Computing the greatest common divisor of two integers up to $n$ bits using Euclid's algorithm has a complexity $O(n^2)$ and there are faster modern methods (see chapter 4 of \cite{BS96}).

\section{Computing the Carmichael function with an order finding oracle}
\label{algo}
Once we have an order finding algorithm, we can devise a sampling algorithm to compute $\lambda(N)$. The order of any $a\in \mathbb{Z}_N^*$ is a divisor of $\lambda (N)$ (Theorem \ref{orderdiv}). If we compute a large enough collection of orders $\ord(a_i)$ from random elements $a_i\in \mathbb{Z}_N^*$, their least common multiple will eventually give the correct value for $\lambda(N)$.

Algorithm \ref{QCarmAlg} gives the pseudocode for the method.

\begin{algorithm}[H]
\caption{Algorithm for computing the Carmichael function $\lambda(N)$}
\label{QCarmAlg}
\begin{algorithmic}[1]
\Procedure{Carmichael}{$N$,$K$}\Comment{Computes $\lambda(N)$}
\State $\lambda \gets 1$
\For{TestedElements $\gets$ $1$ to $K$}
\State Choose a random integer $1 < a < N$.
\State Compute $(a,N)$
\If{$(a,N)\neq 1$}

\State \Return factors $N_1=(a,N)$ and $N_2=N/N_1$. 

\ElsIf{$(a,N)=1$}
\State Compute $\ord(a)$
\State $\lambda \gets \lcm(\lambda,\ord(a))$

\EndIf
\EndFor
\State \Return $\lambda$
\EndProcedure
\end{algorithmic}
\end{algorithm}

The algorithm has two phases. First, we draw a random $1<a<N$ and check $(a,N)=1$. If $a$ is a factor of $N$, we can just return the factors $N_1=a$ and $N_2=N/a$ and reduce the problem of computing $\lambda(N)$ into the two easier problems of computing $\lambda(N_1')$ and $\lambda(N_2')$ where $N_1'$ is the largest power of $N_1=(a,N)$ which still divides $N$ and $N_2'=N/N_1'$. Then, computing $\lambda(N)$ reduces to finding $\lambda(N)=\lcm(\lambda(N_1'),\lambda(N_2'))$ from Equation \eqref{Lcoprime}.

In our discussion, we assume the easiest factors have been discarded. For any $N$, we start by stripping it of its smallest factors. The general quantum algorithm is best used only for the remaining, hard to separate factors of $N$. If it were easy to find a complete factorization of $N$, we could just use Eqs. \eqref{LambdaFromFactors1}{--}\eqref{LambdaFromFactors3} to give $N$.

In the most interesting cases, a random $a$ will have a vanishing probability of being a factor of $N$. 

The algorithm consists in obtaining the order $K$ random elements of $\mathbb{Z}_N^*$ and then computing their least common multiple. The value of $K$ which guarantees the result is $\lambda(N)$ with high probability is discussed in the following subsections.

\subsection{Average number of steps}
The algorithm needs to call the order finding oracle multiple times before returning the correct value of $\lambda(N)$. The number of repetitions $K$ that is required to compute the Carmichael function of $N$ with a probability close to 1 can be deduced from the group structure of $\mathbb{Z}_N^*$ and the concept of primitive roots. 

We just need to remind a few basic results before giving a lower bound to the probability of success when computing $\lambda(N)$. The proofs can be found in Chapter 8 of Burton's book \cite{Bur05}.
 
\begin{definition}[Primitive roots]\label{Proots}
An integer $a\in \mathbb{Z}_N^*$ is called a primitive root of $N$ if $\ord(a)=\varphi(N)$.
\end{definition}

\begin{thm}[Existence of primitive roots]\label{Proots2}
$N$ has a primitive root if and only if it is the power of an odd prime $N=p^{\alpha}$ or $N=2,4, 2p^\alpha$.
\end{thm}

In our first screening we will take out all the small factors of $N$. In the following, we assume $N$ is an odd number (all the powers of 2 have been removed).

\begin{thm}[Number of primitive roots]\label{Proots3}
If $N$ has a primitive root, there are exactly $\varphi(\varphi(N))$ primitive roots.
\end{thm}

The probability bounds follow the analysis of a previous quantum primality test \cite{DG18} where, after finding the order of $O(\log\log N)$ elements, we have a probability close to one of finding an element of order $N-1$ if $N$ is prime. Once we find one element with that order, $N$ is known to be prime with certainty. 

For odd integers that are not prime powers, there are no primitive roots. Instead, the concept of primitive root can be generalized to any $N$ by replacing Euler's function by the Carmichael function \cite{Car10}. A primitive $\lambda$-root is an element of $\mathbb{Z}_N^*$ with the maximum possible order.  

\begin{definition}[Primitive $\lambda$-root]\label{PrimLRootDef}
An integer $a\in \mathbb{Z}_N^*$ is called a primitive $\lambda$-root of $N$ if $\ord(a)=\lambda(N)$.
\end{definition}

\begin{thm}[Number of primitive $\lambda$-roots \cite{Car10}]\label{Proots3}
$\mathbb{Z}_N^*$ has exactly $\varphi(\lambda(N))$ primitive $\lambda$-roots.
\end{thm}

Instead of choosing $K$ so that we have a high probability of finding a primitive $\lambda$-root, we will collect enough $a_i\in \mathbb{Z}_N^*$ to guarantee that $\lambda(N)=\lcm(\ord_N(a_1),\ldots, \ord_N(a_K))$.

We will show $K$ must be of the order of $(\log\log N)^2$. The bound rests on the group structure of $\mathbb{Z}_N^*$. The proofs and further discussion for the following results can be found in Galian's book \cite{Gal17}.

\begin{thm}[Fundamental Theorem of Finite Abelian Groups]\label{fundamental}
Every finite Abelian group is a direct product of cyclic groups of prime-power order.
\end{thm}

Any integer $N$ has a unique prime decomposition (fundamental theorem of arithmetic)
\begin{equation}
N=p_1^{\alpha_1}\cdots p_d^{\alpha_d}=\prod_{i=1}^{d}p_i^{\alpha_i},
\end{equation}
where $d=\omega(N)$ is the number of distinct prime factors of $N$ given by the prime omega function $\omega(N)$. Then, 
\begin{equation}
\mathbb{Z}_N^*=\mathbb{Z}_{p_1^{\alpha_1}}^*\times\cdots\times \mathbb{Z}_{p_d^{\alpha_d}}^*. 
\end{equation}
The elements in the resulting finite Abelian group $\mathbb{Z}_N^*$ can be written as a tuple $a=(a_1,\ldots,a_d)$ with elements $a_i\in \mathbb{Z}_{p_i^{\alpha_i}}^*$. There is an isomorphism between these tuples and the elements in $\mathbb{Z}_N^*$. We can express each $a\in\mathbb{Z}_N^*$ explicitly as the solution to a system of equations
\begin{equation}
a\equiv a_i \mod p_i^{\alpha_i},
\end{equation}
with $(p_i^{\alpha_i},p_j^{\alpha_j})=1$ for $i\neq j$. The Chinese Remainder Theorem guarantees there is only one solution $a\in\mathbb{Z}_N^*$.

Once we have a decomposition of $\mathbb{Z}_N^*$ as a direct product of cyclic groups which have primitive roots, we can relate the order of the elements in each of the cyclic groups to the order of the correspondent element in $\mathbb{Z}_N^*$ \cite{Gal17}.

\begin{thm}[Order of an Element in a Direct Product]\label{OrderProduct}
The order of an element $a$ in a direct product of a finite number of finite groups is the least common multiple of the orders of the
components of the element. 
\end{thm}

\begin{corollary}[Order of an element in $\mathbb{Z}_N^*$]\label{OrderProductZN}
For $a \in \mathbb{Z}_N^*=\mathbb{Z}_{p_1^{\alpha_1}}^*\times\cdots\times \mathbb{Z}_{p_d^{\alpha_d}}^*$ such that $a=(a_1,\ldots,a_d)$:
\begin{equation}
\ord_N(a)=\lcm(\ord_{p_1^{\alpha_1}}(a_1),\ldots,\ord_{p_d^{\alpha_d}}(a_d)).
\end{equation} 
\end{corollary}

We assume an odd $N$. If the original $N$ is even, we can just extract all the powers of 2 and write it as $N=2^r N'$. We can compute $\lambda(N')$ for the odd part with the quantum algorithm and later recover $\lambda(N)$ with Eqs. \eqref{LambdaFromFactors1}{--}\eqref{LambdaFromFactors3}. 
  
For an odd $N=\prod_{i=1}^{d}p_i^{\alpha_i}$, 
\begin{equation}
\lambda(N)=\lcm(p_1^{\alpha_1-1}(p_1-1),\ldots,p_d^{\alpha_d-1}(p_d-1)).
\end{equation}
Once we have an element $a \in \mathbb{Z}_N^*$ with $a=(a_1,\ldots,a_d)$ which includes a primitive root $a_i$ so that $\ord_{p_i^{\alpha_i}}(a_i)=\varphi(p_i^{\alpha_i})=p_i^{\alpha_i-1}(p_i-1)$, we have the whole contribution of $Z_{p_i^{\alpha_i}}^*$ to $\lambda(N)$.

The fraction of elements $a \in \mathbb{Z}_N^*$ which have an order revealing a primitive root of $Z_{p_i^{\alpha_i}}^*$ in their tuple decomposition is, at least,
\begin{equation}
\label{fraction}
F_i=\frac{\varphi(\varphi(p_i^{\alpha_i}))}{\varphi(p_i^{\alpha_i})}=\frac{\varphi(p_i^{\alpha_i-1}(p_i-1))}{p_i^{\alpha_i-1}(p_i-1)}.
\end{equation}
This is the fraction of elements $a_i \in Z_{p_i^{\alpha_i}}^*$ which are primitive roots. In the tuple $a=(a_1,\ldots,a_d)$, all the elements $a\in \mathbb{Z}_N^*$ where $a_i$ is a primitive root of the corresponding cyclic group have an order which is a multiple of $\lambda(p_i^{\alpha_i})$, irrespective of the values of the other $a_j$ for $i\neq j$.

We can find a lower bound for each $F_i$ using the lower bound for $M \geq 3$ \cite{RS62}:
\begin{equation}
\label{LowRosserSchonfeld}
\frac{\varphi(M)}{M}\geq \frac{1}{e^{\gamma}\log\log M+\frac{2.50637}{\log\log M}}
\end{equation}
where $\gamma\approx 0.57721$ is the Euler-Mascheroni constant and $e^\gamma\approx 1.781$. 

We can verify
\begin{equation}
\label{3Mbound}
\frac{\varphi(M)}{M}> \frac{1}{3\log\log M}
\end{equation}
for
\begin{equation}
M>e^{e^{\sqrt{\frac{2.50637}{3-1.781}}}} \approx 49.2.
\end{equation}

As long as every $p_i^{\alpha_i-1}(p_i-1)\geq 50$, we can use the bound in Equation \eqref{3Mbound} on each $F_i$ so that
\begin{equation}
F_i>\frac{1}{3\log\log(p_i^{\alpha_i-1}(p_i-1))}\geq \frac{1}{3\log\log N},
\end{equation}
where the last inequality follows from the fact that $N$ is always equal to or larger than its factors making $N\geq p_i^{\alpha_i-1}(p_i-1)=p_i^{\alpha_i}- p_i^{\alpha_i-1}$.

This bound is valid if each prime power in the factor decomposition of $N$ is greater than 50. As a previous step, $N$ should be checked for small factors. In general, the whole computation is faster with an initial search for factors up to the first few million primes, depending on the speed of the computer doing the checking. Once the easy factors $s_i$ are sieved out, $\lambda(N)=\lambda(s_1\cdots s_k N')$ can be computed using the quantum algorithm to find $\lambda(N')$ and the formulas in Equations \eqref{LambdaFromFactors1}{--}\eqref{LambdaFromFactors3}. 

If we repeat the order finding routine for $K$ random $a\in\mathbb{Z}_N^* $ the probability none of them corresponds to a tuple with a primitive root  of $Z_{p_i^{\alpha_i}}^*$ is $(1-F_i)^K$. 

The total probability of finding at least one element $a\in\mathbb{Z}_N^*$ corresponding to a primitive root of each $Z_{p_i^{\alpha_i}}^*$ after $K$ applications of the order finding oracle is
\begin{equation}
P=\prod_{i=1}^d \left(1-\left(1-F_i\right)^K\right)>\left(1-\left(1-\frac{1}{3\log\log N}\right)^K\right)^d.
\end{equation}
This gives a lower bound for the total probability of success for recovering $\lambda(N)$ as the least common multiple of the orders modulo $N$ of the $K$ random elements we test. Notice we can also obtain the correct value of $\lambda(N)$ even if we fail to find an element corresponding to one of the primitive roots. For instance, if we do not find any elements of $Z_{p_1^{\alpha_1}}^*$ with a contribution of $p_1^{\alpha_1-1}(p_1-1)$ to the order, we still recover the correct contribution to the $\lcm$ if we have two elements which contribute to the order with factors $p_1^{\alpha_1-1}$ and $p_1-1$ each. Likewise, there might be repeated factors corresponding to different $Z_{p_i^{\alpha_i}}^*$.

We want $P$ to be arbitrarily close to $1$. We must choose a $K$ so that 
\begin{equation}
\label{smallerror}
d\left(1-\frac{1}{3\log\log N}\right)^K\ll 1
\end{equation} 
making 
\begin{equation}
P\approx 1-d\left(1-\frac{1}{3\log\log N}\right)^K \approx 1.
\end{equation} 
 
If we take logarithms on both sides of Eq. \eqref{smallerror}, we have a condition on $K$
\begin{equation}
\log d+ K \left(1-\frac{1}{3\log\log N}\right)\ll 0.
\end{equation}
As $N$ grows, $\frac{1}{3\log\log N}\ll 1$ and we can bound the number of repetitions as
\begin{equation}
\log d- K \frac{1}{3\log \log N}\ll 0
\end{equation} 
which becomes
\begin{equation}
K \gg 3\log \log N \log d.
\end{equation} 

The smallest possible prime factor of $N<2^n$ is $p=2$ and we can use $n$ as an upper bound for $d$, giving a number of repetitions
\begin{equation}
K \gg 3\log n \log d=(\log n)^2.
\end{equation} 

Up to a constant, the number of repetitions is of the order of $O((\log n)^2)$. We can choose a constant an order of magnitude above 3, like 50.

\subsection{Total complexity}
In total, we obtain $\lambda(N)$ with high probability after $O((\log n)^2)$ calls to the order finding oracle, which has a complexity $O((\log n )n^3)$.

The whole procedure needs of the order of $O((\log n )^3n^3)$ quantum operations giving a complexity essentially cubic in the number of bits of $N$, which could be reduced to $O(\log\log n (\log n)^4 n^2)$ in the large $N$ limit where fast multiplication is more efficient.

In the classical part of the algorithm we need to compute the $\gcd$ and the $\lcm$ of different elements. The complexity of the Euclidean algorithm for the $\gcd$ is quadratic and the $lcm(x,y)=\frac{|xy|}{\gcd(x,y)}$ uses the gcd, multiplication and division with also an asymptotic quadratic complexity. 

The cost is similar to an alternative approach using Shor's algorithm. In the algorithm for the Carmichael function, we have found a bound of $O((\log n)^2)$ calls to order finding. Finding two factors of $N$ on a quantum computer requires $O(\log n)$ calls to the order finding oracle. In the factoring approach to computing $\lambda(N)$, we have to look at the total number of prime factors, including multiplicities. This number is given by the function $\Omega(N)\geq \omega(N)$, which is always, at least, as large as the number of distinct prime factors $\omega(N)$. For a total factorization we need to repeat Shor's algorithm for a maximum of $n-1$ times until we recover all the factors \cite{CL97} for a worst case complexity bounded by $O(n\log n)$ calls to the order oracle. 

In practice, both factoring with Shor's algorithm and the proposed Carmichael function algorithm can be much simpler. For instance, semiprimes of the form $N=pq$ (the product of two primes) are particularly interesting because of their use in the RSA cryptosystem. For semiprimes $\omega(N)=\Omega(N)=2$. 

Additionally, as $N$ grows, both $\omega(N)$ and $\Omega(N)$ have an average value, or normal order, $\log\log(N)=\log(n)$ (Hardy-Ramanujan theorem \cite{HR17}). In fact, 
\begin{equation}
\frac{\omega(N)-\log\log(N)}{\sqrt{\log\log N}}
\end{equation}
follows, essentially, a Gaussian normal distribution for large integers (Erd\H{o}s-Kac theorem \cite{EK40}).

The expected complexity to find $\lambda(N)$ will be $O(\log\log n \log n)$ calls to the order finding oracle for our $\lcm$ approach and $O(\log(n)^2)$ for the factoring method.
 
One advantage of the method proposed in this paper is that the results are accumulative. We could start with a smaller number of repetitions and, even if during verification we find our value $\lambda'(N)$ is not the true Carmichael function of $N$, we can still compute $\lcm(\lambda'(N),\lambda''(N))$ combining the result $\lambda''(N)$ of a new attempt. The result is the same as taking together the combined total number of elements $a \in \mathbb{Z}_N^*$ used in both runs. 

Finally, there are multiple tradeoffs in factoring. The quantum part of the method can be substantially reduced at the cost of more sophisticated classical processing. For instance, in most typical cases, Shor's algorithm can be reduced to a single call of the order finding oracle \cite{Eke21}. Similar methods can probably be used in the direct computation of $\lambda(N)$.

As the typical complexity is usually smaller than the worst case bound of $O((\log n)^2)$ calls to the order finding oracle and the result can be quickly verified (see next Section), the Carmichael function algorithm can be run in successive rounds until a valid result is found. While this introduces a verification overhead in the intermediate stages, it can reduce the average complexity.  

\section{Verification}
\label{verification}
Once we computed a candidate for $\lambda(N)$ from the $\lcm$ of the order of $K$ elements, we need to verify it is actually the right value.

We assume the order finding oracle is correct and we recover the minimum value for which the powers of $a$ repeat themselves. With quantum order finding, the probability this value is indeed the smallest one is close to 1.

Now we have discarded errors by excess (obtaining a multiple of $\lambda(N)$), we need to check for errors by defect where the $K$ elements we took are not enough to get the full value of $\lambda(N)$ and we get, instead, $l|\lambda(N)$.

In this case, it will be easy to find an element $a$ such that $a^l\not\equiv 1 \mod N$ which serves as a witness $l$ is not the correct value for the Carmichael function. 

The elements of $a \in  \mathbb{Z}_N^*$ with $a^l\equiv 1 \mod N$ are a subgroup of $\mathbb{Z}_N^*$. For $l\neq \lambda(N)$ the subgroup must be stricly smaller than $\mathbb{Z}_N^*$. At least some elements, the primitive $\lambda$-roots, have an order $\ord_N(a)=\lambda(N)$ and, for them, $a^l\not\equiv 1 \mod N$. There are $\varphi(\lambda(N))\geq 1$ from these $\lambda$-roots (Theorem \ref{Proots3}). 

Using Lagrange's Theorem (Theorem \ref{LagrangeTh}), we see the number of elements in the $a^l\equiv 1 \mod N$ subgroup must divide $|\mathbb{Z}_N^*|=\varphi(N)$. The smallest possible divisor is 2, meaning that at least half of the $a \in  \mathbb{Z}_N^*$ will fail to satisfy $a^l\equiv 1 \mod N$. 

After running the proposed algorithm, we just need to draw $k$ random elements of $\mathbb{Z}_N^*$ to verify $\lambda(N)$ is the actual value of the Carmichael funtion with a probability greater than $2^{-k}$.

\section{Some quantum improvements}
\label{qimprovements}
The general method has some interesting possibilities for fine tuning in a quantum setting.

One direct saving comes from using the quantum period finding algorithm to avoid computing the $\lcm$ of the order of multiple elements at the cost of increasing the size of the input. 

We start from the initial state
\begin{equation}
\frac{1}{N}\sum_{r=0}^{N^2-1}\ket{r}\ket{a_1}\ket{a_2}\cdots\ket{a_K}
\end{equation}
with $K$ registers, one for every different random element $a_i\in \mathbb{Z}_N^*$. Like in the basic quantum order finding algorithm, we compute the superposition of modular exponentials
\begin{eqnarray}
&\frac{1}{N}&\sum_{r=0}^{N^2-1}\ket{r}\ket{a_1^r \mod N}\ket{a_2^r \mod N}\cdots\ket{a_K^r \mod N} \nonumber\\
=&\frac{1}{N}&\sum_{r=0}^{N^2-1}\ket{r}\ket{A}.
\end{eqnarray}
Each of the $K$ registers repeats its value with a period $\ord(a_i)$. The complete state $\ket{A}$ only repeats when all the periods synchronize. The period of this larger value is the least common multiple of the periods of each individual register starting in the state $\ket{a_i}$.

This reduces the number of repetitions to a single call to the order oracle at the cost of increasing the second register from $n$ to $O(n(\log n)^2)$ qubits. While interesting, this is probably not practical. The classical part of the algorithm computing the $\lcm$ is not so costly and increasing the number of qubits is usually more difficult. 

We can also perform a quantum verification of the result using a generalization of Fermat's and Euler Theorems.

\begin{thm}[Maximal generalization of Fermat's theorem \cite{Sin66}]\label{GenFer}
For any integers $a$ and $N=p_1^{\alpha_1}\cdots p_d^{\alpha_d}$
\begin{equation}
a^s\equiv a^{s+t} \mod N
\end{equation}
for any $t\geq s$ such that $\lambda(N)|(t-s)$ with $s\geq \max(\alpha_1,\ldots,\alpha_d)$.
\end{thm}
The smallest possible factor of $N$ is 2, so the maximum possible $\alpha_i$ is $\log_2(N)$. If we choose $s>\log_2(N)$, for instance $n$, and $t=s+\lambda(N)$, $a^s\equiv a^{s+t} \mod N$ is true for any $0 \leq a\leq N$, even for values $a_i\not\in \mathbb{Z}_N^*$ such as the factors of $N$ when $N$ is composite. 

With this theorem, we can verify $\lambda(N)$ in three steps. First, we prepare a uniform superposition
\begin{equation}
\frac{1}{\sqrt{N}}\sum_{a=0}^{N-1}\ket{a},
\end{equation}
which can be done by taking the $n$-qubit QFT for an initial $\ket{0}$ state. Now we add an ancillary qubit to store the difference between $a^{s} \mod N$ and $a^{s+t} \mod N$ for our candidate $t$ for $\lambda(N)$ to obtain the state
\begin{equation}
\frac{1}{\sqrt{N}}\sum_{a=0}^{N-1}\ket{a}\ket{a^s-a^{s+t} \mod N}.
\end{equation}
We measure the second register. If $t=\lambda(N)$, the state is always $\ket{0}$. If the result is not $\ket{0}$, $t\neq \lambda(N)$ and we can measure the first register to get a witness. The result is either an element $a \in \mathbb{Z}_N^*$ such that $a^t\not\equiv 1 \mod N$ or an integer $a\not\in \mathbb{Z}_N^*$, which should also follow Theorem \ref{GenFer}. In the second case, $gcd(a,N)$ gives us a factor of $N$. If we obtain a non-trivial factor, we can reduce the complexity of the problem by dividing it into smaller pieces with Eqs. (\ref{LambdaFromFactors1}-\ref{LambdaFromFactors3}).

Up to this point, there is no advantage with respect to the classical method of choosing a random $1<a<N$ and verifying $a^s\equiv a^{s+t} \mod N$. If $t\neq \lambda(N)$, the probability of finding a witness classically and the probability of reducing the state of the superposition to a new state where the second register is not $\ket{0}$ is the same. 

However, the quantum method gives a second chance to verify the result. Imagine $t\neq \lambda(N)$ and that, when we measure the second register, we find the state $\ket{0}$. The resulting superposition will have less than $N-\varphi(N)/2$ states. For at least half the elements in $\mathbb{Z}_N^*$, $a^t\not\equiv 1 \mod N$ and $a^s\not\equiv a^{s+t} \mod N$.

We can now take the QFT of this new superposition and measure the result. For this state, the probability of finding the $\ket{0}$ state is equal to the number of elements of the superposition divided by $N$. If $t=\lambda(N)$ the QFT of the resulting uniform superposition becomes $\ket{0}$ with certainty. If $t\neq \lambda(N)$, we might obtain a state $\ket{x}$ with $x \neq 0$ which serves as a witness that our candidate for the Carmichael function is wrong. 

Finding $\ket{0}$ on measurement is inconclusive, as it was in the classical test finding an element $a$ such that $a^t\equiv 1 \mod N$. What we gain with this second chance is an additional shot at detecting a false value. The probability of finding a state different from $\ket{0}$ is the same as finding a witness during the first measurement. With this trick, we can cut the verifications of the classical strategy by a factor of two at the cost that any detection in this second stage will not give us a witness $a$ for which we can verify classically $a^s\not\equiv a^{s+t} \mod N$.

After $k$ repetitions of this procedure without finding a classical or a quantum witness, we be satisfied that $t=\lambda(N)$ with a probability of error smaller than $\left(\frac{\varphi(N)}{2N}\right)^{2k}$. If we have done an initial screening for small factors, for large $N$, $\frac{\varphi(N)}{N}\to 1$ or we can find factors efficiently by random sampling. This gives a bound to the maximum probability of error which is close to $2^{-2k}$.

\section{Discussion}
\label{discussion}
We have seen quantum computers can be used to compute the Carmichael function for any integer $N$ directly, without the need to factor $N$. The algorithm takes $O((\log n)^2)$ calls to the order oracle and can be run on a quantum computer using $O((\log n )^3n^3)$ elementary gates or $O(\log\log n (\log n)^4 n^2)$ operations for fast multiplication.

There are a few applications for the algorithm. First, it gives a primality test. An integer $N$ is a prime if and only if $\lambda(N)=N-1$, which can be deduced from the converse of Fermat's theorem \cite{Luc78}. This complements a previous quantum primality test searching for primitive roots of order $N-1$ \cite{DG18}.

This property also allows us to identify Carmichael numbers: composite integers $N$ for which all the $a\in \mathbb{Z}_N^*$ satisfy $a^{N-1}\equiv 1 \mod N$ \cite{Car12}. If they obey the condition Fermat's theorem, $\lambda(N)|N-1$. Any integer with a Carmichael function different from, but a factor of $N-1$ is a Carmichael number. 

Finally, the algorithm provides a direct attack against the RSA cryptosystem \cite {RSA78} without factoring $N$ explicitly. For the ciphertext $C=M^e \mod N$ corresponding to a message $M$, a public key $e$ and the public modulus $N$, we can compute the modular inverse $d\equiv e^{-1} \mod \lambda(N)$ such that $ed\equiv 1 \mod N$ and $M^{ed}\equiv M^{k \lambda(N)+1}\equiv M \mod N$, which gives the any valid message $M$ from its ciphertext.

\section*{Acknowledgements} 
This work has been funded by the Spanish Government and FEDER grant PID2020-119418GB-I00 (MICINN) and Junta de Castilla y Le\'on (project VA296P18).
 

\newcommand{\noopsort}[1]{} \newcommand{\printfirst}[2]{#1}
  \newcommand{\singleletter}[1]{#1} \newcommand{\switchargs}[2]{#2#1}
\end{document}